\documentclass{article}
\usepackage{spconf,amsmath,graphicx}
\usepackage{url,lipsum}
\usepackage{tabularx}
\usepackage{amsmath,amssymb}
\usepackage{xpatch,xcolor}
\usepackage{makecell, booktabs}
\usepackage{algorithm}
\usepackage{algpseudocode}
\newcolumntype{Y}{>{\centering\arraybackslash}X}
\usepackage{xpatch,xcolor}
\usepackage{cite}
\usepackage[subtle]{savetrees}

\usepackage{hyperref}

\hypersetup{
    colorlinks=true,
    linkcolor=purple,
    filecolor=magenta,      
    urlcolor=purple,
}

\newcommand{\newpara}[1]{\vspace{8pt}\noindent\textbf{#1}}
\def\BigRoman{\uppercase\expandafter{\romannumeral\number\count 255 }}
\def\Romannumeral{\afterassignment\BigRoman\count255=}

\title{Multi-scale speaker embedding-based \\graph attention networks for speaker diarisation}

\name{\parbox{\linewidth}{\centering Youngki Kwon, Hee-Soo Heo, Jee-weon Jung, You Jin Kim,  Bong-Jin Lee, Joon Son Chung}}
\address{
Naver Corporation, South Korea\\
}
\begin{document}
\maketitle
\ninept

\begin{abstract}
The objective of this work is effective speaker diarisation using multi-scale speaker embeddings.
Typically, there is a trade-off between the ability to recognise short speaker segments and the discriminative power of the embedding, according to the segment length used for embedding extraction. 
To this end, recent works have proposed the use of multi-scale embeddings where segments with varying lengths are used. 
However, the scores are combined using a weighted summation scheme where the weights are fixed after the training phase, whereas the importance of segment lengths can differ with in a single session.

To address this issue, we present three key contributions in this paper:
(1) we propose graph attention networks for multi-scale speaker diarisation;
(2) we design scale indicators to utilise scale information of each embedding;
(3) we adapt the attention-based aggregation to utilise a pre-computed affinity matrix from multi-scale embeddings.

We demonstrate the effectiveness of our method in various datasets where the speaker confusion which constitutes the primary metric drops over 10\% in average relative compared to the baseline.

\end{abstract}


\section{Introduction}
\label{sec:intro}


Speaker diarisation is a task of partitioning audio clips (i.e., sessions) into speaker homogeneous segments, essentially solving the problem of {\em ``who spoke when''}. 
It is an important pre-processing step for many speech-related tasks, such as transcribing meetings or movie scripts.

Recent work on speaker diarisation can be divided into two strands. The first is end-to-end based methods~\cite{fujita19_interspeech,fujita2019end,horiguchi20_interspeech}, which have received increasing attention recently. 
Although these works have shown promise in limited environments, they have been reported not to generalise well to real-world conditions. 
The second strand employs multiple conventional modules, comprising a multi-stage pipeline. 
While the stages differ by the systems, they typically consist of speech activity detection, speaker embedding extraction, and clustering. 
Researches in this strand are focused on optimising the pipeline.

In particular, the quality of speaker embedding plays a crucial role in the performance of speaker diarisation systems. 
In general, speaker embeddings produced from longer segment are more discriminative. 
However, since these embeddings are more likely to contain multiple speakers, they are more vulnerable to rapid speaker transitions.
Deriving speaker embeddings from a short duration mitigates this issue, but, their discriminativeness is known to be weaker.
Because of this trade-off, recent literature in speaker diarisation adopts a window of 1.5 seconds for extracting speaker embeddings.

A few works have addressed this issue with using {\em multi-scale} speaker embeddings where the different scales refers to the duration of audio for embedding extraction~\cite{park2021multi,wang2021bytedance}. 
Different scales are combined using weights assigned to each scale. 
Various techniques have been introduced to derive these weights.
However, none of these works involve adaptive mechanisms that can assign weights dynamically. 
Thus, once the weights are fixed, it is applied to entire sessions. 
Because of the limitation of the existing framework, speaker embeddings can be only compared with embeddings where the scale is identical.
This work proposes to address these limitations. 

In this study, we propose several techniques on top of the existing multi-scale speaker embedding framework. 
Specifically, we introduce a graph attention network (GAT)~\cite{PetarVelickovic2018GRAPHNETWORKS} in place of weighted summation when merging multi-scale speaker embeddings.
The graph attention networks can model non-Euclidean relations between speaker embeddings with different scales where each embedding is set as a node. 

Our approach can compare embeddings with different scales where the weights are dynamically assigned via the attention mechanism. 
It can even assign different weights within a single session which improves the flexibility. 
A novel {\em scale indicator}, inspired by positional encoding~\cite{parmar2018image, vaswani2017attention}, is also proposed.
We additionally extend the attention-based aggregation method, introduced in our previous work~\cite{kwon2021adapting}, by modifying the mechanisms towards multi-scale. 
We expect that it would be a novel application of multi-scale embeddings in the speaker diarisation.

We validate our approach using a wide range of datasets, DIHARD {\Romannumeral 1}, {\Romannumeral 2} and {\Romannumeral 3} test sets, as well as the VoxConverse dataset. 
Consistent improvement in all scenarios is observed, demonstrating the effectiveness of our approach.




\section{Multi-scale speaker diarisation pipeline}
\label{sec:pipeline}

\begin{figure}[t!]
    \centering
    \includegraphics[width=\columnwidth]{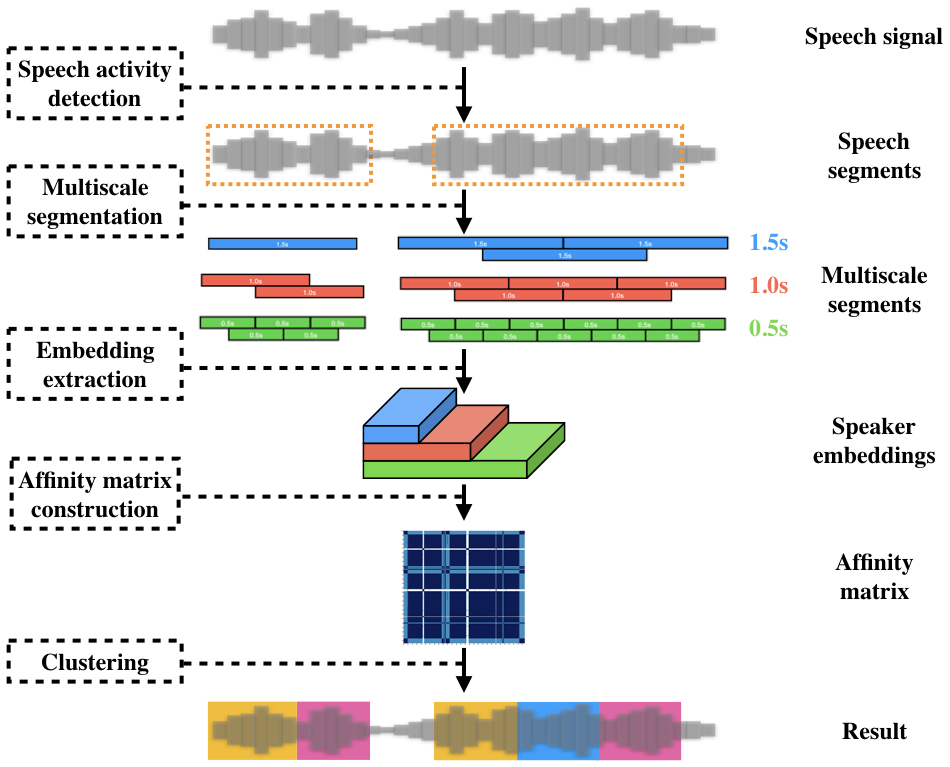}
    \caption{Multi-scale speaker diarisation pipeline.}
    \label{fig:overview_pipeline}
\end{figure}

In this section, we describe the overall process pipeline of speaker diarisation using multi-scale speaker embeddings. 
The overall structure is illustrated in Figure~\ref{fig:overview_pipeline}. 
Speech segments are extracted from the input audio via SAD. Uniform segmentation is performed for all speech segments extracted using different time scales. Speaker embeddings are extracted for each scale segment. We then construct an affinity matrix based on these speaker embeddings. The resulting affinity matrix is used for clustering, the final step in the pipeline. The details of each step are explained through subsections.

\subsection{Speech activity detection}
The speech activity detection (SAD) module is located at the front of the speaker diarisation pipeline. 
It performs segmentation regardless of speaker identity. 
The output of the SAD module is denoted using onsets and offsets of each segment where overlaps and speaker changes can be involved. 
We employ a convolutional recurrent neural network-based SAD module which has an identical architecture to that of our previous work~\cite{jung2021three}. 
We further apply a sliding window on the module's output to decide onsets and offsets where an onset is determined when more than 70\% of a window is predicted as speech and vice versa for an offset. 

\subsection{Multi-scale segmentation}

Using the SAD module's output, we split each segment into groups of shorter segments where the shorter segments in each group have an equal duration (i.e., scale) and different groups have different scales. 
This multi-scale segmentation uses four hyper-parameters: number of scales, segment scale defined by window size and shift size, base-scale which determines the unit of clustering and labeling, and segment mapping criterion which defines how to map between segments from different scales.
We employ the method introduced in \cite{park2021multi}. 
We use three different scales. 
The window sizes for each scale are 0.5s, 1.0s, and 1.5s, and the shift sizes are 0.25s, 0.25s, and 0.16s, respectively. 
We use the scale of 0.5s window size as a base-scale. 
For segment mapping, we select the segment with the closest midpoint as suggested in \cite{park2021multi}.

\subsection{Speaker embedding extraction}
Speaker embeddings are derived for each and every segment divided via the multi-scale segmentation. 
We build our speaker embedding extraction module on top of that we used in our previous work~\cite{kwon2021adapting}, which is a variant of \cite{kwon2020ins}. 
Three additional techniques are additionally employed in the training phase. 

First, we use {\em mixup}~\cite{zhang2017mixup} which generates augmented training samples using a weighted summation of two different speakers' utterances and a corresponding soft-label. 
Second, we propose a new augmentation technique by connecting two different speakers' utterances also trained with soft-labels.
Lastly, each mini-batch is constructed using one of the scales among the utilised multi-scales.

The former two techniques aim to counteract towards segments that include overlaps or speaker changes whereas the last technique is applied to facilitate multi-scale speaker embedding extraction.


\subsection{Clustering}
\label{ssec:Clustering}
We assign a speaker label for each segment via spectral clustering\cite{von2007tutorial,ning2006spectral} using the speaker embeddings. 
Spectral clustering has been widely adopted in speaker segmentation pipelines. 
It is considered as a manifold-based clustering technique, where results highly depend on the quality of the affinity matrix. 
Elements of the affinity matrix represent the similarity between two speech segments.

Various methods can be used to measure the similarity when constructing the affinity matrix $\mathbf{M}$ using multi-scale speaker embeddings. 
An existing work applies weighted summation of cosine similarities of each scale segment pair introduced in \cite{park2021multi}.


\begin{equation}
    \begin{aligned}
  m_{ij} = Sim(S_i, S_j) = w_{0.5} * cos(\mathbf{e}_{i,0.5}, \mathbf{e}_{j,0.5})\\
  \:\:\:+ \, w_{1.0} * cos(\mathbf{e}_{i,1.0}, \mathbf{e}_{j,1.0})\\
  \:\:\:+ \, w_{1.5} * cos(\mathbf{e}_{i,1.5}, \mathbf{e}_{j,1.5}), 
    \end{aligned}
\end{equation}

\noindent where $S_i$ is the $i'th$ segment, $\mathbf{e}_{i,s} \in \mathbb{R}^d$ is the embedding extracted $i'th$ segment at scale $s$, and $w_s \in \mathbb{R}$ is weight factor for scale $s$.


This method is used as the baseline of our work.



\begin{figure}[t!]
    \centering
    \includegraphics[width=\columnwidth]{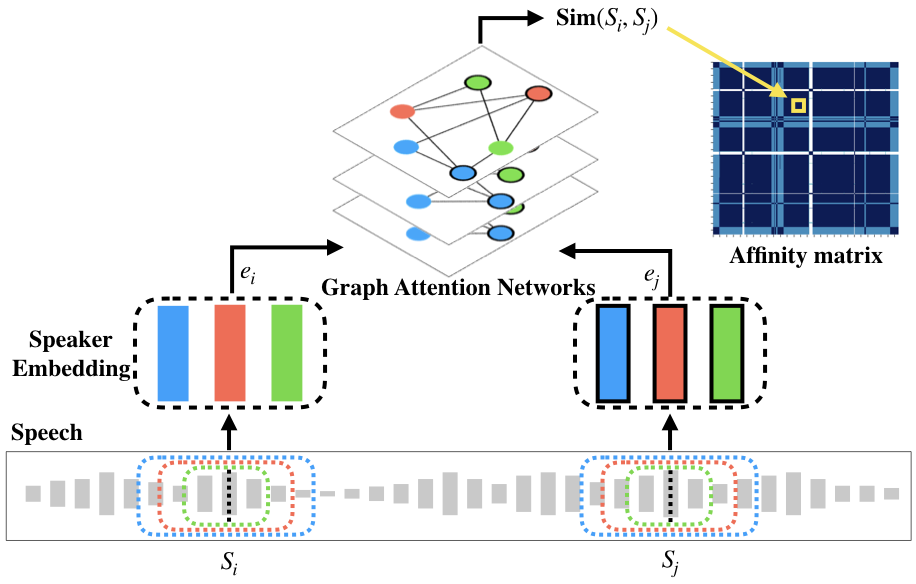}
    \caption{Overview of proposed GAT-based similarity measure.}
    \label{fig:overview}
\end{figure}

\section{Proposed GAT module}
\label{sec:proposed}
In this section, we address our novel GAT module that inputs multi-scale speaker embeddings and outputs an integrated, single value composing the affinity matrix.
We first introduce our variant of the original GAT architecture, which has been successfully adopted to speaker verification and spoofing detection in our previous works~\cite{jung2021graph,tak2021end,tak2021graph,jung2021aasist}.
Then, we address how we derive an integrated affinity matrix using multi-scale speaker embeddings.

The purpose of the GAT used in this work is to calculate similarity $Sim(S_i, S_j)$ where $S_i$ and $S_j$ are the $i'th$ and $j'th$ segment, respectively.

\begin{equation}
    \begin{aligned}
        Sim(S_i, S_j) = GAT(\mathcal{G}).
    \end{aligned} 
\end{equation}

First, we construct a graph $\mathcal{G}$ using multiple embeddings from two segments as: 

\begin{equation}
    \begin{aligned}
        \mathcal{G} = \{MS_{i}, MS_{j}\},
    \end{aligned} 
\end{equation}

\noindent where $MS_{i} = \{ \mathbf{e}_{i,0.5}, \mathbf{e}_{i,1.0}, \mathbf{e}_{i,1.5} \}$ is the set of multi-scale embeddings extracted from the $i'th$ segment.

However, since the existing GAT architecture cannot utilise the scale information of each embedding, a scale indicator is proposed. 
The scale indicator involves three vectors, each corresponding to one among three scales (0.5s, 1.0s, and 1.5s). 
These vectors are learnable parameters like the rest of the parameters within the model. 
We integrate scale indicators into the GAT model using an element-wise addition between each node and its corresponding scale indicator.
Through this mechanism, we leverage our prior knowledge on which scale the speaker embedding (node) is extracted from.
The dimension of the scale indicator is identical to that of input nodes to make element-wise addition applicable.
This technique is inspired by the positional encoding of the Transformer~\cite{parmar2018image, vaswani2017attention}. 
With the scale indicator, the node representation $\mathbf{h}_u$ is defined as:

\begin{equation}
    \begin{aligned}
        \mathbf{h}_u = \mathbf{e}_{i,s} \oplus \mathbf{l}_{s},
    \end{aligned} 
\end{equation}

\noindent where $\mathbf{h}_u$ refers the node representation of node $u$ and vector $\mathbf{l}_{s}$ is scale indicator for each scale $s$.

Since it is difficult to pre-define the relationship between each node like in the previous research, it is assumed that all nodes are connected (i.e., all possible edges exist), including self-connections.
Instead, the relations between nodes are defined by an attention mechanism assigning a weight for each edge using learnable parameters. 
In particular, the attention value $\alpha_{u,v}$ between node $u$ and $v$ is defined as:

\begin{equation}
    \alpha_{u,v}=\frac{exp(g(\mathbf{h}_{u}, \mathbf{h}_{v}))}{\sum_{k \in \mathcal{G} }  exp(g(\mathbf{h}_{u}, \mathbf{h}_{k}))  }, 
\end{equation}

\noindent where the attention function $g(\cdot, \cdot)$ is defined as following:

\[
    g(\mathbf{h}_{u}, \mathbf{h}_{v})= \\ 
\begin{cases}
    (\mathbf{h}_{u} \otimes \mathbf{h}_{v})\mathbf{w}_{1},& \text{if } u \in \mathcal{U}(v)\\
    (\mathbf{h}_{u} \otimes \mathbf{h}_{v})\mathbf{w}_{2},              & \text{otherwise}
\end{cases}
\]

\noindent where $\mathbf{w}_{1}$ and $\mathbf{w}_{2}$ are the weight vectors for each case, $\otimes$ is the element-wise multiplication, and $\mathcal{U}(v)$ is a set of nodes from the same segment.
Different from the previous approach which only compares pairs of the same scale, the attention function defined above allows comparisons between different scales because it considers all possible node pairs.
In addition, by using two attention parameters ($\mathbf{w}_{1}$ and $\mathbf{w}_{2}$), it is possible to perform aggregation between discriminative nodes within the same segment while comparing nodes from the different segments.

Finally, the similarity between the two segments calculated through the above steps is used to construct the affinity matrix.
After constructing the affinity matrix, clustering can be performed as described in the previous section~\ref{ssec:Clustering}.
The node configuration and operation of the GAT is common to that of the GAT structure applied to speaker verification and spoofing detection~\cite{jung2021graph,tak2021end,tak2021graph}.

\section{Attention-based feature enhancement}

Our previous work~\cite{kwon2021adapting} refines single-scale speaker embeddings before constructing the affinity matrix, referred to as {\em attention-based embedding aggregation} (AA).
Although this technique brings performance improvement in a stable manner, it cannot be applied to multi-scale speaker embeddings. 
We thus adapt AA towards multi-scale framework as described in Algorithm~\ref{alg:ModifiedAA}. 

The proposed algorithm utilises the multi-scale affinity matrix $\mathbf{M}$, described in Section~\ref{sec:pipeline} and a set of speaker embeddings. 
The speaker embeddings are used to construct $\mathbf{C}$.
However, affinity matrix ($\mathbf{C}$) is constructed by speaker embeddings in base scale (0.5s) segments may not be as good due to its embedding quality.
To overcome this issue, we match the shape by replacing base-scale speaker embeddings with larger-scale speaker embeddings when constructing $\mathbf{C}$.  
For example, a base-scale embedding that represents 2.0s - 2.5s is replaced with a large-scale speaker embedding that represents 1.5s - 3.0s. 
The objective is to make $\mathbf{C}$ in Algorithm~\ref{alg:ModifiedAA} credible.



We use each affinity matrix to construct an attention map. $\mathbf{A}_1$ is constructed from the affinity matrix $\mathbf{M}$ and $\mathbf{A}_2$ is constructed from the affinity matrix $\mathbf{C}$ (lines 5-6). We then create an attention map $\mathbf{A}$ by adding the weights of the two attention maps (line 7).
The weight values of both matrices change with each step.
We first construct the attention map $\mathbf{A}$ so that $\mathbf{A}_1$ has a higher weight, and then we gradually build up the attention map $\mathbf{A}$ so that $\mathbf{A}_2$ has a higher weight relative to $\mathbf{A}_1$.
 


\begin{algorithm}[!t]
    \caption{Multi-scale attention based feature enhancement}
    \label{alg:ModifiedAA}
    \begin{algorithmic}[1]
    \State \textbf{Input:} Base-scale speaker embeddings $\mathbf{X} \in \mathbb{R}^{\textit{L} \times 256}$, multi-scale affinity matrix $\mathbf{M} \in \mathbb{R}^{L \times L}$
    \State \textbf{Hyper-parameters:} Number of repetition $N$, Temperature value $\tau$
    \For {$i=0,1,\ldots,N-1$}
		\State Construct affinity matrix $\mathbf{C}|c_{i,j}=cos(\textbf{X}_{i},\textbf{X}_{j})$
		\State $\mathbf{A}_1$ = softmax($\mathbf{M}$ * $\tau$)
		\State $\mathbf{A}_2$ = softmax($\mathbf{C}$* $\tau$)
		\State $\mathbf{A}$ = (($N$ - $i$) * $\mathbf{A}_1$ + $i$ * $\mathbf{A}_2$) / $N$
		\State $\mathbf{X}$ = dot($\mathbf{A}$, $\mathbf{X}$)
	\EndFor
\end{algorithmic}
\end{algorithm}

\section{Experiments and results}
We conduct experiments to evaluate the performance of multi-scale diarisation pipelines on various datasets: the test set of the first, second, third DIHARD challenge\cite{ryant2018first,Ryant2019TheBaselines,ryant2020third} and VoxConverse\cite{Chung2020SpotWild}.

\begin{table}[!t]
	\centering
	\small
	\caption{
	  Results on the DIHARD {\Romannumeral 1}, {\Romannumeral 2}, {\Romannumeral 3} and VoxConverse test sets.  (FA: false alarm, MS: miss, SC: speaker confusion, \textbf{lower is better for all four metrics}). Except for GAT + AA, all configuration with AA use the original version of AA~\cite{kwon2021adapting}.  
	}
	\begin{tabularx}{\columnwidth}{lYYYY}
    \Xhline{1pt}
	 Configuration & DER & FA & MS & SC\\ 
	 \Xhline{1pt}
	 \multicolumn{5} {c} {\bf{DIHARD {\Romannumeral 1}} }\\
     \hline
    \Xhline{1pt}
	 Baseline (0.5s) & 31.40 & 0.0 & 8.71 & 22.69 \\
	 Baseline (1.0s) & 25.46 & 0.0 & 8.71 & 16.75 \\
	 Baseline (1.5s) & 24.60 & 0.0 & 8.71 & 15.88 \\
	 Cosine Fusion & 25.89 & 0.0 & 8.71 & 17.18 \\
	 GAT & 23.26 & 0.0 & 8.71 & 14.55 \\
	 \hline
	 Baseline (0.5s) + AA & 31.68 & 0.0 & 8.71 & 22.97 \\
	 Baseline (1.0s) + AA & 22.29 & 0.0 & 8.71 & 13.58 \\
	 Baseline (1.5s) + AA & 20.54 & 0.0 & 8.71 & 11.83 \\
	 Cosine Fusion + AA & 23.44 & 0.0 & 8.71 & 14.72 \\
	 GAT + AA & {\bf{19.00}} & 0.0 & 8.71 & 10.29 \\
    \Xhline{1pt}
    \multicolumn{5} {c} {\bf{DIHARD {\Romannumeral 2}} }\\
     \Xhline{1pt}
     Baseline (0.5s) & 35.61 & 0.0 & 9.69 & 25.93 \\
	 Baseline (1.0s) & 27.60 & 0.0 & 9.69 & 17.91 \\
	 Baseline (1.5s) & 27.87 & 0.0 & 9.69 & 18.18 \\
	 Cosine Fusion & 28.13 & 0.0 & 9.69 & 18.44 \\
	 GAT & 22.85 & 0.0 & 9.69 & 13.16 \\
	 \hline
	 Baseline (0.5s) + AA & 35.28 & 0.0 & 9.69 & 25.59 \\
	 Baseline (1.0s) + AA & 22.77 & 0.0 & 9.69 & 13.08 \\
	 Baseline (1.5s) + AA & 21.47 & 0.0 & 9.69 & 11.78 \\
	 Cosine Fusion + AA & 23.10 & 0.0 & 9.69 & 13.41 \\
	 GAT + AA & \bf{19.80} & 0.0 & 9.69 & 10.12 \\
    \Xhline{1pt}
    \multicolumn{5} {c} {\bf{DIHARD {\Romannumeral 3}} }\\
    \Xhline{1pt}
     Baseline (0.5s) & 25.80 & 0.0 & 9.52 & 16.28 \\
	 Baseline (1.0s) & 20.08 & 0.0 & 9.52 & 10.56 \\
	 Baseline (1.5s) & 19.93 & 0.0 & 9.52 & 10.40 \\
	 Cosine Fusion & 20.48 & 0.0 & 9.52 & 10.96 \\
	 GAT & 18.32 & 0.0 & 9.52 & 8.79 \\
	 \hline
	 Baseline (0.5s) + AA & 25.01 & 0.0 & 9.52 & 15.49 \\
	 Baseline (1.0s) + AA & 17.50 & 0.0 & 9.52 & 7.98 \\
	 Baseline (1.5s) + AA & {\bf{17.04}} & 0.0 & 9.52 & 7.52 \\
	 Cosine Fusion + AA & 18.25 & 0.0 & 9.52 & 8.73 \\
	 GAT + AA & 17.35 & 0.0 & 9.52 & 7.83 \\
    \Xhline{1pt}
    \multicolumn{5} {c} {\bf{VoxConverse} }\\
    \Xhline{1pt}
     Baseline (0.5s) & 28.94 & 1.38 & 3.29 & 24.27 \\
     Baseline (1.0s) & 23.99 & 1.38 & 3.29 & 19.32 \\
     Baseline (1.5s) & 24.81 & 1.38 & 3.29 & 20.12 \\
     Cosine Fusion & 22.98 & 1.38 & 3.29 & 18.31 \\
	 GAT & 18.62 & 1.38 & 3.29 & 13.94 \\
	 \hline
     Baseline (0.5s) + AA & 28.76 & 1.38 & 3.29 & 24.08 \\
	 Baseline (1.0s) + AA & 21.64 & 1.38 & 3.29 & 16.96 \\
	 Baseline (1.5s) + AA & 14.88 & 1.38 & 3.29 & 10.21 \\
	 Cosine Fusion + AA & 18.95 & 1.38 & 3.29 & 14.27 \\
	 GAT + AA & {\bf{12.78}} & 1.38 & 3.29 & 8.1 \\
    \Xhline{1pt}
	\end{tabularx}
	\label{tab:results}
\end{table}

\label{sec:ExpAndResult}
\subsection{Evaluation protocol}
\label{ssec:evaluation_protocol}
We use the DER (Diarisation Error Rate) as a primary metric. The DER is consists of three components: False alarm (FA, speech in prediction but not in reference), Missed speech (MS, speech in reference but not in prediction), Speaker confusion (SC, speech assigned to wrong speaker ID). We use the {\tt dscore}\footnote{https://github.com/nryant/dscore} to measure the performance. 

In order to test VoxConverse, we conduct experiments under the condition of using system SAD, and in this case, we use 250ms as the collar. For the DIHARD dataset, experiments are performed under the condition which is provided with a reference SAD, and we use 0 ms as the collar.

\subsection{Implementation details}
\label{ssec:details}

\subsubsection{Clustering}\label{sssec:clustering_details}
The clustering stage of our diarisation pipeline requires a threshold for eigenvalues. 
We tune the threshold for each dataset based on empirical evaluations; the values are 48, 38, 48, and 80 for DIHARD {\Romannumeral 1}, DIHARD {\Romannumeral 2}, DIHARD {\Romannumeral 3}, and VoxConverse, respectively.

\subsubsection{Attention-based feature enhancement}
AA requires two hyperparameters: the temperature value and the number of repetition. We use 0.30 as the temperature value for all datasets. Use a different value for each dataset for the number of repetition. Actual values are 10 for DIHARD {\Romannumeral 1}, 20 for DIHARD {\Romannumeral 2}, 10 for DIHARD {\Romannumeral 3}, and 15 for VoxConverse.

\subsubsection{Graph attention networks}
\newpara{Creating training data and label.}
As with the speaker verification datasets~\cite{Chung18b,Nagrani17}, we construct a training dataset containing speaker pairs and labels.
These pairs are extracted from RTTM files of the development data of speaker diarisation datasets.
It uses speakers from RTTM to form a speaker pair. For each RTTM, all combinations of two speakers are selected and then multi-scale speech segments with the same midpoint are extracted.
We use the DIHARD {\Romannumeral 1}, {\Romannumeral 2}, {\Romannumeral 3} development sets~\cite{ryant2018first,Ryant2019TheBaselines,ryant2020third}, VoxConverse development sets~\cite{Chung2020SpotWild}, ICSI datasets~\cite{janin2003icsi}, AMI datasets~\cite{carletta2005ami}, and internal conversation datasets as source data.
We also conduct data augmentations following recipes: applying room impulse response~\cite{Ko2017ARecognition,Povey2011TheToolkit}, channel corruption using FFMPEG, masking frequency bin higher than 4k to simulate narrow-band signal.

\newpara{Training protocol.}
In this dataset configuration, positive pairs are scarce. 
Thus, without further interference, GAT would tend to ignore positive pairs and output all inputs as negative.
To reduce this tendency, we oversample positive pairs with an imbalanced dataset sampler~\footnote{https://github.com/ufoym/imbalanced-dataset-sampler} and construct a mini-batch of evenly composed pairs of both types.

Binary cross-entropy loss is used to train the GAT model. The model is trained for 50 epochs using a mini-batch size of 50. While the GAT model is being trained, the speaker embedding model is fixed. We train the model using Adam optimizer~\cite{Kingma2015adam}. A learning rate of 0.0001 is used as the initial value, after which it is adjusted through the cosine annealing scheduler~\cite{loshchilov2016sgdr}.

\subsection{Baselines}
We test two types of baselines: single-scale diarisation systems and multi-scale diarisation systems. The baseline for a multi-scale diarisation system is described in Section ~\ref{sec:pipeline}. Cosine Fusion in Table ~\ref{tab:results} represents this system. A single-scale diarisation system that has the same structure in a multi-scale baseline except using only a single-scale segment. Baseline (0.5s), Baseline (1.0s), and Baseline (1.5s) in Table ~\ref{tab:results} represent these systems. And we also test those mentioned systems in combination with AA.

\subsection{Results analysis}
\label{ssec:analysis}

Table ~\ref{tab:results} shows the diarisation results in DIHARD {\Romannumeral 1}, {\Romannumeral 2}, {\Romannumeral 3}, and VoxConverse. 
All configurations are evaluated with and without the proposed AA technique where top five rows denote performance without AA.
A single-scale system shows better numbers when using a larger time scale segment. 
Cosine Fusion is slightly underperforms compared to the best single-scale systems. 

The multi-scale pipeline with the GAT module shows significant performance improvements relative to the baselines. This performance improvement can be seen on all datasets that we tested. Compared to the single-scale baseline of the best performance, the speaker confusion is improved by an average of 19.56\%, and compared to the multi-scale baseline (Cosine Fusion), there is a performance improvement of 21.90\%.

The above-mentioned trend can be confirmed even when feature enhancement (AA) is used. Compared to the single-scale baseline, an average improvement of 10.91\% is shown and compared to the multi-scale baseline, there is a performance improvement of 27.04\%.

Also, the modified feature enhancement has an effect. There is an average increase of 26.30\% (GAT vs. GAT + AA).

\section{Conclusion}
\label{sec:conclusion}

We proposed a graph attention network for multi-scale speaker diarisation. This module enabled the construction of an affinity matrix by using segments of varying lengths to compute the similarity between the two segments. For a more robust similarity calculation, we designed scale indicators that provide scale information for each multi-scale speaker embedding. We also revamped the previous feature enhancement technique to take advantage of the pre-computed affinity matrix, which was constructed from multi-scale embeddings. We evaluated our proposed method on various datasets and demonstrated consistent performance improvement across a range of datasets.

\clearpage
\bibliographystyle{IEEEbib}
\bibliography{shortstrings,refs}

\end{document}